\documentclass[
]{ceurart}

\usepackage{listings}
\begin{document}

%%
%% Rights management information. You may not change this.
\copyrightyear{2024}
\copyrightclause{Copyright for this paper by its authors.
  Use permitted under Creative Commons License Attribution 4.0
  International (CC BY 4.0).}

%%
%% This command is for the conference information
\conference{CAMLIS'24: Conference on Applied Machine Learning for Information Security,
  October 24--25, 2024, Arlington, VA}

%%
%% The "title" command
\title{Applying Action Masking and Curriculum Learning
Techniques to Improve Data Efficiency and Overall
Performance in Operational Technology Cyber
Security using Reinforcement Learning}

% \tnotemark[1]
% \tnotetext[1]{Use this to make an author note if needed}

%%
%% The "author" command and its associated commands are used to define
%% the authors and their affiliations.
\author[1]{Alec Wilson}[%
orcid=0009-0003-9181-9766,
email=Alec.Wilson@uk.bmt.org,
]
% \cormark[1]
% \fnmark[1]
\address[1]{BMT, London, UK}

\author[2]{William Holmes}[%
email=William@adsp.ai,
]
% \fnmark[1]
\address[2]{ADSP, London, UK}

\author[1]{Ryan Menzies}[%
orcid=0009-0003-9646-196X,
email=Ryan.Menzies@uk.bmt.org,
]

\author[1]{Kez {Smithson Whitehead}}[%
orcid=0009-0005-2286-2140,
email=Kez.SmithsonWhitehead@uk.bmt.org,
]

%% Footnotes
% \cortext[1]{Corresponding author.}
% \fntext[1]{These authors contributed equally.}

%%
%% The abstract is a short summary of the work to be presented in the
%% article.
\begin{abstract}
  In previous work, the IPMSRL environment (Integrated Platform Management System Reinforcement Learning environment) was developed with the aim of training defensive RL agents in a simulator representing a subset of an IPMS on a maritime vessel under a cyber-attack. This paper extends the use of IPMSRL to enhance realism including the additional dynamics of false positive alerts and alert delay. Applying curriculum learning, in the most difficult environment tested, resulted in an episode reward mean increasing from a baseline result of -2.791 to -0.569. Applying action masking, in the most difficult environment tested, resulted in an episode reward mean increasing from a baseline result of -2.791 to -0.743. Importantly, this level of performance was reached in less than 1 million timesteps, which was far more data efficient than vanilla PPO which reached a lower level of performance after 2.5 million timesteps. The training method which resulted in the highest level of performance observed in this paper was a combination of the application of curriculum learning and action masking, with a mean episode reward of 0.137. This paper also introduces a basic hardcoded defensive agent encoding a representation of cyber security best practice, which provides context to the episode reward mean figures reached by the RL agents. The hardcoded agent managed an episode reward mean of -1.895. This paper therefore shows that applications of curriculum learning and action masking, both independently and in tandem, present a way to overcome the complex real-world dynamics that are present in operational technology cyber security threat remediation.
\end{abstract}

%%
%% Keywords. The author(s) should pick words that accurately describe
%% the work being presented. Separate the keywords with commas.

%%
%% This command processes the author and affiliation and title
%% information and builds the first part of the formatted document.
\maketitle

\section{Introduction}
In previous work, the IPMSRL environment \cite{wilson_multi-agent_2023} was developed with the aim of training defensive RL agents in a simulator representing a subset of an IPMS on a maritime vessel under a cyber-attack. This work explores the impact of changing the difficulty of the simulator through the manipulation of values representing real world dynamics, e.g. False negative rate of alerts. \par
RL agents will often be significantly limited if they are not exposed to the environment in which they are intended to be deployed. This also translates to when trained agents are deployed in a real scenario. Consequently, the environment needs to replicate the real scenario as closely as possible. \par
This paper extends the use of IPMSRL to enhance realism including the additional dynamics of false positive alerts and alert delay. Additionally, three configurations of the environment of varying degrees of difficulty are defined and tested to understand the different levels of performance a trained RL agent can reach. \par
This paper also applies curriculum learning and action masking as ways to mitigate the increased levels of difficulty, showing that using these techniques is data efficient and leads to higher mean episode reward. \par
Curriculum learning alone is shown to increase mean episode reward \cite{narvekar_curriculum_2020}, \cite{skinner_reinforcement_1958}. Action masking alone is shown to similarly improve mean episode reward \cite{vinyals_starcraft_2017}, \cite{openai_dota_2019}, \cite{huang_closer_2022}. Action masking also has the additional benefit of significantly faster training and the ability to constrain an agent’s available action space to a set that meets user-defined criteria. \par
Finally, both curriculum learning and action masking are applied together. This training method resulted in the highest level of mean episode reward observed in this paper.

\section{Background}
\subsection{Reinforcement Learning}
\begin{figure}[hbt!]
    \centering
    \includegraphics{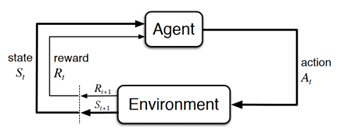}
    \caption{Reinforcement Learning Architecture \cite{sutton_reinforcement_2018}.}
    \label{fig:rl_architecture}
\end{figure}
RL is a training method where an agent learns to interact with an environment to complete a task. The environment is a Markov Decision Process (MDP) and consists of a state space, action space, reward function, and a transition model \cite{sutton_reinforcement_2018}. At each timestep an agent will take an action on the environment. The agent will then receive a reward from the reward function and the updated state of the environment. The goal of RL is for the agent to learn to maximise the reward signal. We choose to use RL over other AI methods as it allowed the agent to learn without the need for existing datasets.

\subsection{IPMSRL}
\begin{figure}[hbt!]
    \centering
    \includegraphics[width=1\linewidth]{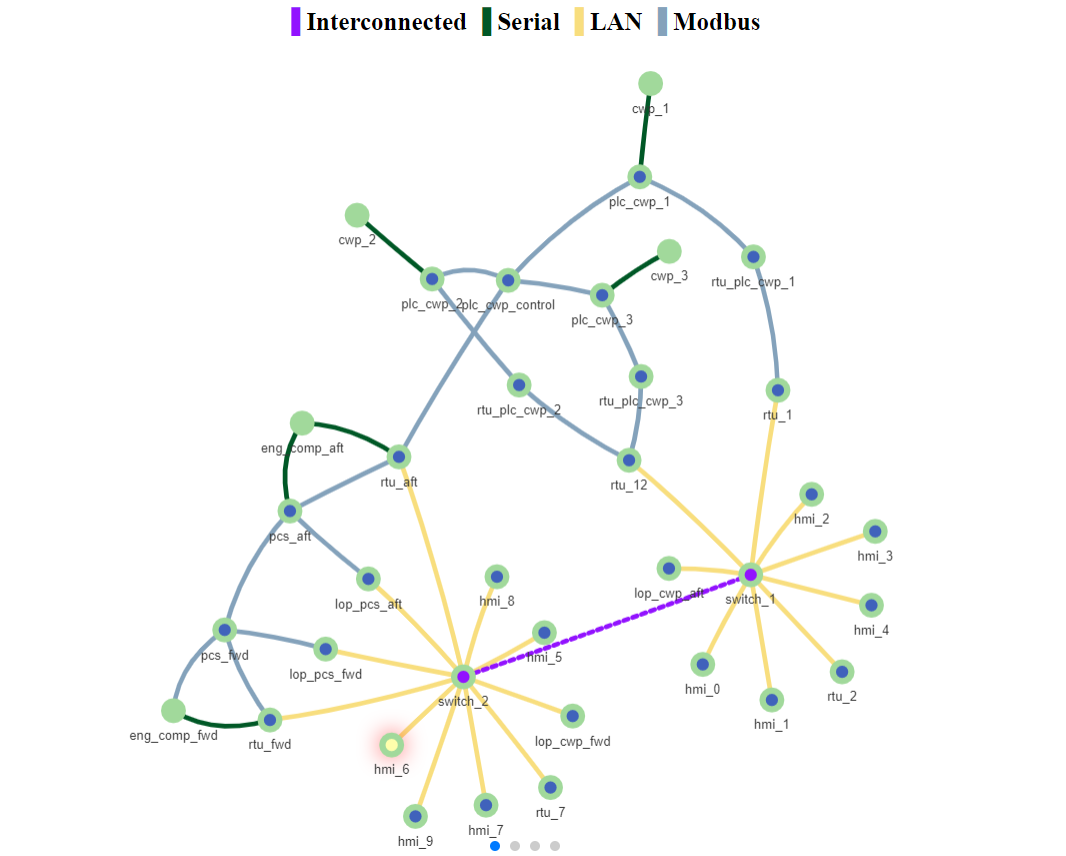}
    \caption{IPMSRL Environment \cite{wilson_multi-agent_2023}.}
    \label{fig:ipmsrl_env}
\end{figure}

IPMSRL \cite{wilson_multi-agent_2023} is a Gymnasium-based \cite{towers_gymnasium_nodate} Reinforcement Learning (RL) environment that simulates an Integrated Platform Management System (IPMS) on a vessel under a cyber-attack. An IPMS controls and monitors many ship systems across propulsion, power, steering, stability, auxiliary and ancillary systems. To achieve this, IPMS utilises a distributed control system architecture that facilitates interfaces with sensors, equipment, plants, software-based control systems and network-based data \cite{wilson_multi-agent_2023}. A physical representation of the IPMSRL environment has been shown in Figure \ref{fig:ipmsrl_env}.
The configuration of IPMSRL used in this paper controls a subset of these systems and focuses on the propulsion and chilled water system. The defending RL agent receives intrusion detection alerts on components following simulated infection by a cyber attacker. These alerts are based upon the MITRE ATT\&CK ICS framework\footnote{© 2024 The MITRE Corporation. This work is reproduced and distributed with the permission of The MITRE Corporation.} \cite{corporation_ics_2023}. These alerts are then passed onto the defending agent through the observation space which the agent uses to represent the environment state, $S_{t}$, as shown on Figure \ref{fig:rl_architecture}. Subsequently, the agent chooses a discrete action, $A_{t}$ {contain, eradicate, recover, or wait}, for a given node. Contain, eradicate, and recover represents an action space modified from NIST SP-800-61 guidance, which was adapted to an Operational Technology (OT) scenario \cite{wilson_multi-agent_2023}, \cite{nist_nist_2012}. An instantaneous timestep, $t$, takes place and the environment produces a reward, $R_{t+1}$, and updated state $S_{t+1}$. \par
This sequence is then repeated until all critical nodes are compromised which would result in a negative reward or all infections are completely removed which would result in a positive reward. We also implemented an early stopping criterion of 50 timesteps.

\subsection{PPO} 
Proximal Policy Optimisation (PPO) is one of the current state-of-the-art algorithms used within RL and Multi-Agent Reinforcement Learning (MARL) \cite{schulman_proximal_2017}, \cite{yu_surprising_2022}. PPO applies the policy gradient method within the actor-critic architecture. The algorithm was chosen as it is robust to hyperparameter tuning and has been shown to be performant in previous work \cite{wilson_multi-agent_2023}, \cite{schulman_proximal_2017}. The actor, or namely the policy network, chooses the action given an observation. The critic, or namely the value network, produces an estimate of the sum of future rewards given the action from the policy network and state. This can be simplified to the actor chooses an action and the critic assesses its quality. The hyperparameters and architecture used in our experiments have been provided in the appendix. 

\subsection{Curriculum Learning and Action Masking}
Curriculum Learning and Action Masking are two popular guided RL methods to address data efficiency concerns in RL \cite{narvekar_curriculum_2020}, \cite{vinyals_starcraft_2017}, \cite{openai_dota_2019}, \cite{eser_guided_2023}. This paper explores both forms of guided RL for the IPMSRL cyber security environment \cite{wilson_multi-agent_2023}, initially individually and subsequently in combination, which we show further improves performance. 

\subsection{Curriculum Learning}
Curriculum Learning (CL) in RL is the process of increasing the difficulty of a task by periodically shaping aspects of the MDP throughout training \cite{narvekar_curriculum_2020}, \cite{skinner_reinforcement_1958}. This type of guided RL is often implemented by altering aspects of the environment such as the complexity of the transition function to increase the difficulty of optimising the reward function \cite{vinyals_starcraft_2017}. As a result, CL can be considered a form of transfer learning and has been shown to be beneficial for sim-to-real applications \cite{diprasetya_sim--real_2024}.\par
CL has been shown to reduce learning time and improve performance of the trained agent for applications including robotics and games \cite{bengio_curriculum_2009}, \cite{silver_mastering_2016}. In this paper, we explore if CL offers similar benefits in the domain of cyber security by testing it on the IPMSRL environment \cite{wilson_multi-agent_2023}. We introduce three stages to the curriculum (Easy, Medium and Hard) where difficulty is defined as the uncertainty in the transition function. We show CL outperforms training directly on the Hard environment configuration.

\subsection{Action Masking}
Action Masking (AM) is a guided RL method which allows the integration of additional human knowledge into the learning process \cite{eser_guided_2023}. AM limits the range of actions by introducing guardrails which can prevent undesirable actions from being chosen by the agent. These guardrails require action space shaping which has similar disadvantages to reward shaping, such as increased manual set up time and susceptibility to human bias \cite{kanervisto_action_2020}. However, a key advantage of guardrails is that they can provide both increased data efficiency and user-defined constraints which are both essential considerations for cyber security applications \cite{thakur_investigation_2015}. \par
Action masking limits the set of actions that can be executed per timestep based on the current environment state. In the context of masking for discrete spaces, this is the process of reducing the choice of actions so that undesirable or impossible actions are unavailable to the agent. Specifically, this is achieved by setting the probabilities of selecting the undesirable actions to zero or near zero during stochastic learning \cite{vinyals_starcraft_2017}. \par
Previous work has shown action masking can simplify learning for the agent and result in reduced training times in video games including StarCraft II and DOTA 2 \cite{vinyals_starcraft_2017}, \cite{openai_dota_2019}, \cite{huang_closer_2022}. The focus of the work was primarily to address the high dimensionality of the action space as opposed to implementing safety critical constraints, but existing work \cite{muller_safe_2022} has shown how action masking can be applied in RL to improve safety for traffic-based applications.\par
The data efficiency benefits of action masking in cyber security applications with discrete action spaces were explored in this paper. Specifically, we mask invalid actions on the IPMSRL environment and show that the learning process provides a higher average return, as the agent focuses on learning only from the valid set of actions. This helped prevent the agent from wasting time exploring trajectories and taking undesirable actions that would not be applicable for deployment. In future work, the use of action masking could be further extended to restrict available actions based on safety-critical criteria.\par
We show below how action masking can also be applied to improve the realism of the IPMSRL simulation, including full details of how masking was applied in our experiments. For example, in the real world if the agent has sent a command to contain a node, then the user would likely have to wait for this process to complete before sending a different command to the given node. 

\subsection{Combined Curriculum and Action Masking }
AM and CL often aim to achieve the same benefits of safer learning and reduced training time. We show below how these methods can be combined to improve performance over each method individually. Existing work has shown how automatic action masking can be used as a type of CL to alter the action space \cite{yasutomi_automatic_2023}. In contrast, we show action masking can be applied to the action space and used with vanilla CL as a method to address the uncertainty in the environment’s transition function. 
\section{Experiments}
\subsection{Environment Difficulty Configurations}
In a real scenario, Security Information and Event Management (SIEM) systems are used to give increased visibility of an OT system and flag any potential malicious activity. SIEMs are not perfect and suffer from False Positives (FP) and False Negatives (FN). The IPMSRL environment used in this paper has the additional feature of FP alerts. \par
Three defined environment difficulties were explored: easy, medium and hard: Table 1 shows the values chosen for each parameter and difficulty. The easy configuration has no FP or FN alerts, no alert delays and a 100\% action success rate. The medium and hard configurations add difficulty to the environment by amending the FP, FN, action success probabilities and the alert delay. In previous work, it was demonstrated that as alert and action success probabilities decreased (FN increasing), a reduction to the agent’s performance was observed \cite{wilson_multi-agent_2023}. \par

% Please add the following required packages to your document preamble:
% \usepackage[table,xcdraw]{xcolor}
% Beamer presentation requires \usepackage{colortbl} instead of \usepackage[table,xcdraw]{xcolor}
\begin{table}[hbt!]
\centering
\caption{Easy, medium and hard configurations}
% Please add the following required packages to your document preamble:
% \usepackage[table,xcdraw]{xcolor}
% Beamer presentation requires \usepackage{colortbl} instead of \usepackage[table,xcdraw]{xcolor}
\begin{tabular}{lllll}
\rowcolor[HTML]{00B6DD} 
Parameter Name & Parameter Description & Easy & Medium & Hard \\
\rowcolor[HTML]{CBE5F2} 
\begin{tabular}[c]{@{}l@{}}Alert False \\ Positives \\ Probability\end{tabular} & \begin{tabular}[c]{@{}l@{}}The probability of an \\ alert false positive \\ going off on each node \\ per step\end{tabular} & 0 & 0.01 & 0.03 \\
\rowcolor[HTML]{E7F3F9} 
\begin{tabular}[c]{@{}l@{}}Alert Success \\ Probability\end{tabular} & \begin{tabular}[c]{@{}l@{}}The probability an alert \\ is successfully detected \\ on each infected node \\ per step\end{tabular} & 1 & 0.9 & 0.75 \\
\rowcolor[HTML]{CBE5F2} 
\begin{tabular}[c]{@{}l@{}}Alert Delays \\ (Dependent on \\ MITRE Step)\end{tabular} & \begin{tabular}[c]{@{}l@{}}The delay in timesteps \\ between the node being \\ infected/increasing \\ infection level and the \\ alert going off\end{tabular} & \begin{tabular}[c]{@{}l@{}}(MITRE 1-4): 0\\    \\ (MITRE 5-8): 0\\    \\ (MITRE 9-12): 0\end{tabular} & \begin{tabular}[c]{@{}l@{}}(MITRE 1-4): 1\\    \\ (MITRE 5-8): 0\\    \\ (MITRE 9-12): 0\end{tabular} & \begin{tabular}[c]{@{}l@{}}(MITRE 1-4): 2\\    \\ (MITRE 5-8): 1\\    \\ (MITRE 9-12): 0\end{tabular} \\
\rowcolor[HTML]{E7F3F9} 
\begin{tabular}[c]{@{}l@{}}Action success \\ probability\end{tabular} & \begin{tabular}[c]{@{}l@{}}Probability the \\ defender action \\ is successful\end{tabular} & 1 & 0.9 & 0.75
\end{tabular}
\end{table}

The alerts in the IPMSRL environment are based on the MITRE ATT\&CK ICS Tactics \cite{corporation_ics_2023}, with each tactic representing a different type of alert. These tactics are given a tactic level to represent them with the first tactic listed given a level of 1, with the following tactics increasing in level incrementally. The tactics are: Initial Access, Execution, Persistence, Privilege Escalation, Evasion, Discovery, Lateral Movement, Collection, Command and Control, Inhibit Response Function, Impair Process Control and Impact.\par
The alert delay parameter breaks the 12 tactics into 3 sections, displayed in Table \ref{tab:env_configs}. In the easy difficulty environment there is no delay present, in the medium difficulty environment there is a small delay to alerts earliest in the attack on a given node, the hard difficulty environment has a larger delay for the earliest stage tactic alerts and a small delay for tactics level 5-8. The rationale behind earlier stage tactics receiving a larger delay is that, generally, these tactics are likely to have a higher threshold which will need to be reached before malicious activity on a given node or network can be detected and reported as an alert, as the activity associated with these tactics is harder to differentiate from user activity. Therefore, as the environment increases in difficulty, the alert delay increases towards a more realistic scenario. \par
It is necessary to point out that although IPMSRL has added support for more realistic and complex dynamics, it is still an abstract representation of an IPMS and an attack on this system. The different difficulties of environment add this realism to test the performance of different training approaches and algorithms, but further work needs to be completed on IPMSRL before it can be considered representative of a real-world system.\par
All of the experiments conducted in this paper use a PPO \cite{schulman_proximal_2017} based agent, trained for 2.5 million timesteps over 4 seeds with a 95\% confidence interval (CI). The hyperparameters are available in the appendix. \par

\subsection{Hardcoded Defender}
A basic hardcoded defender was designed with logic developed alongside a cyber security expert. This hardcoded defender is not intended to be perfect, but in the absence using cyber security experts to act as the defender remediating the threat within the IPMSRL environment, the hardcoded defender deploys solid logic based on the NIST SP-600-61 guidance \cite{nist_nist_2012}. The hardcoded defender is therefore able to provide us context as to what level of mean episode reward represents a “good” performance with validated logic. The hardcoded defender takes one action per timestep, as an RL agent does in all of the experiments explored in this paper. The defensive hardcoded agent uses the following logic to determine which action to take next: 
\begin{enumerate}
    \item Contain infectable node that is connected to a critical node, with an alert.
    \item Recover offline critical node.
    \item Contain infectable node with an alert.
    \item Eradicate infectable node that is connected to critical node, with an alert.
    \item Eradicate infectable node with an alert.
    \item Recover infectable node that is in a contained state.
    \item Wait.
\end{enumerate}
The logic follows the fundamental idea of sequentially containing, eradicating and then recovering a node, with some modifications to prioritise recovering offline critical nodes and remediating nodes that are ‘closer’ to critical nodes first.\par
The hardcoded defender was able to reach a mean episode reward over 10,000 episodes of 0.988 in an easy environment, 0.883 in a medium difficulty environment and -1.895 in a hard environment configuration. Figure \ref{fig:cl_am_final} shows the comparison of the best performing RL defenders and the hardcoded defender.\par
\subsection{Baseline Results for Vanilla PPO}
The baseline results of a single agent acting in the IPMSRL environment with varying degrees of difficulty are shown in Figure 2. In the easy configuration, the agent reached near optimum performance after 1 million timesteps, winning every episode and achieving an extremely high episode reward mean value of 0.977. For the medium configuration, the agent reached an episode reward mean of 0.104 and in the hard configuration, the agent struggled to perform well, resulting in an episode reward mean of -2.791. These baseline results for vanilla PPO are all below that of the episode reward mean achieved by the hardcoded defender.\par
\begin{figure}[hbt!]
    \centering
    \includegraphics[width=0.75\linewidth]{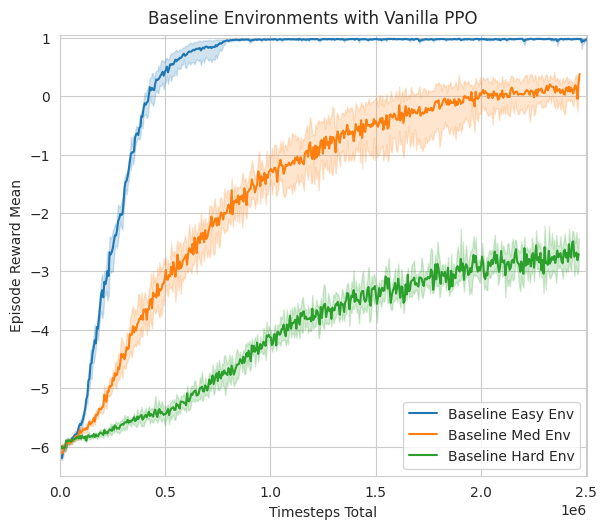}
    \caption{Baseline for different environment configurations.}
    \label{fig:baseline}
\end{figure}

The results in Figure 3 show that there are clear challenges for the agent to learn an optimum policy using a standard training process with PPO \cite{schulman_proximal_2017} when the difficulty of the environment configuration increases. For this reason, we explored methods which enabled the data efficiency and overall performance of the agent to improve significantly. 

\subsection{Curriculum Learning}
As discussed in the background section, CL is a training method which, in this implementation, allows the agent to initially explore simpler tasks where the agent is expected to explore more effectively, before the task is changed to a more difficult one. This enables the incremental increase in the difficulty of the environment during the agent’s training.\par
The tasks can be changed at pre-defined points based on either standard RLlib training metrics e.g. mean episode reward, or custom metrics which are calculated in the IPMSRL environment e.g. win rate. This is implemented through RLlib’s TaskSettableEnv API \cite{rllib_cl_2023}.\par
For all the CL results reported in this paper, the number of total timesteps was used to change tasks in each training sample. The number of total timesteps was chosen to be the point at which the agent, at the current task difficulty, had reached a ‘stable’ level of performance e.g. the agent’s mean episode reward had plateaued. Therefore, for different training curricula, tasks are changed at different points depending on how quickly the agent can reach a stable level of performance. During this experiment, the tasks were changed at 850k timesteps and 1.7m timesteps. Approximately a third of the total training was completed at each task difficulty, once the task’s training had converged. The light blue lines in Figure \ref{fig:curriculum_learning} show the points at which the tasks were changed, and performance subsequently dropping sharply.
\begin{figure}[hbt!]
    \centering
    \includegraphics[width=0.75\linewidth]{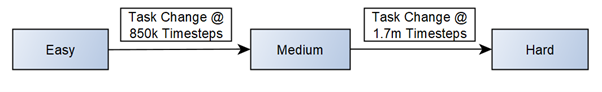}
    \caption{Curriculum for CL Experiment.}
    \label{fig:cl_experiment}
\end{figure}
Figure \ref{fig:cl_experiment} shows the curriculum used in this experiment. Figure \ref{fig:curriculum_learning} compares the results of an agent trained on the hard environment configuration (Figure \ref{fig:baseline}), with an agent trained via a curriculum of easy, medium then hard difficulty configurations. \par
In Figure \ref{fig:curriculum_learning}, it is clearly shown that the agent trained via CL can reach a significantly higher level of performance within the training time when compared to the baseline of “vanilla” training of a PPO agent in the hard environment configuration. The episode reward mean reached was -0.569 in comparison to the baseline of -2.791. The CL agent also significantly outperforms the hardcoded defensive agent’s performance in a hard environment, where it scored an episode reward mean of -1.895. This behaviour is expected and supports the related literature’s conclusions \cite{bengio_curriculum_2009}, \cite{silver_mastering_2016} and the intuitive premise that by leveraging the previous learning of transferable skills in simpler environments to more complex environments, an agent will perform better than attempting to learn this behaviour from scratch in a far more complex environment. 
\begin{figure}[hbt!]
    \centering
    \includegraphics[width=0.75\linewidth]{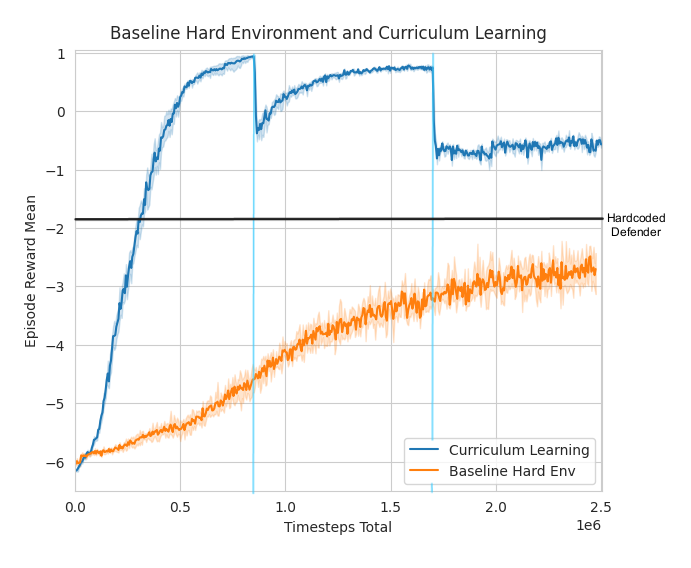}
    \caption{Baseline Hard Environment and Curriculum Learning. The changes in task are outlined by the light blue lines for the CL experiments.}
    \label{fig:curriculum_learning}
\end{figure}
\subsection{Action Masking}
The implementation of action masking encompasses two primary components: modifications within the environment and the development of custom models that incorporate action masking. The implementation was adapted from examples provided within RLlib \cite{liang2018rllibabstractionsdistributedreinforcement}. \par
Action masking utilises a binary array (mask) to identify permissible actions at each step.  For this purpose, a specialised class is created, which, upon initialisation, reads a configuration file to ascertain the applicable mask conditions. This class features a method that is called at every environment step to evaluate which actions should be masked based on the current state of the environment. The observation space, structured as a dictionary, facilitates the mask's transfer to the model, by including “action mask” and “observations” keys. The “action mask” is a 1D binary array indicating valid and invalid actions, while “observations” provide a conventional representation of the environment's state.\par
For the custom model incorporating discrete action masking a PyTorch implementation is used, which is compatible with RLlib and functional for DQN and Policy-Gradient style algorithms, such as PPO \cite{schulman_proximal_2017}. This model initialises an internal fully connected network that processes solely the observation component of the observation space. In the forward pass, it computes unmasked logits by feeding the observation component through this internal network. Additionally, the action mask is transformed into an infinite mask, setting valid actions to 0 and invalid actions to a large negative value (effectively negative infinity), ensuring invalid actions are highly unlikely to be selected post-softmax application. The logits derived from the internal model are then added to this infinite mask, enabling the selection of a non-masked action. This process for action masking can be integrated with other custom models, such as a centralised critic model, allowing for both centralised critic functionality and action masking.\par
All instances of action masking reported in this paper used the masking conditions based on input from a cyber security expert to reflect realistic cyber defence constraints and logic. These masks are deliberately simple to avoid over-engineering and unnecessary bias. The implemented masks are: \par
\begin{itemize}
    \item If there is no alert on an infectable node, mask the contain and eradicate action on that node.
    \item If the infectable node is contained, mask the contain action on that node.
    \item If the infectable node is not contained, mask the eradicate and recover action on that node.
    \item If an action is already in progress on an infectable node, mask all actions on that node.
    \item If a critical node is online, mask the recover action on that node.
\end{itemize}

\begin{figure}[hbt!]
    \centering
    \includegraphics[width=0.75\linewidth]{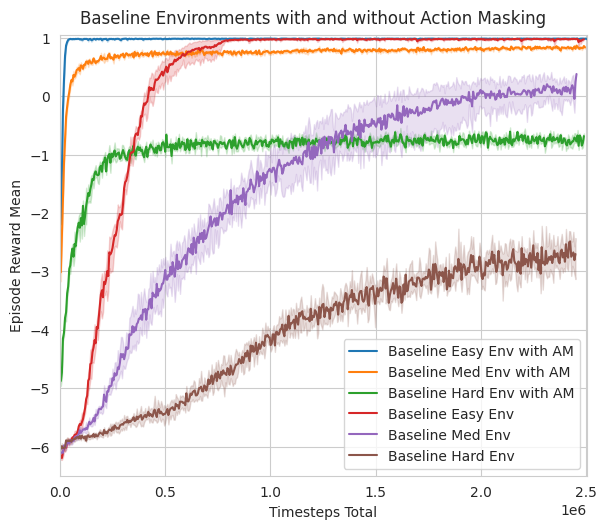}
    \caption{Baseline Environments with Action Masking.}
    \label{fig:AM_baseline}
\end{figure}

Figure \ref{fig:AM_baseline} shows the baseline results of the easy, medium and hard environment configurations when action masking is applied, compared to the baseline results without action masking. In all instances, both a dramatic improvement in data efficiency and overall performance compared to vanilla PPO can be seen. In the easy configuration, for all seeds, the agent reached an optimum level of performance in less than 100k timesteps, winning every episode, resulting in an episode reward mean of 0.977. In the medium difficulty configuration, the agent reached a good level of performance, an episode reward mean of 0.816. \par

\begin{figure}[hbt!]
    \centering
    \includegraphics[width=0.75\linewidth]{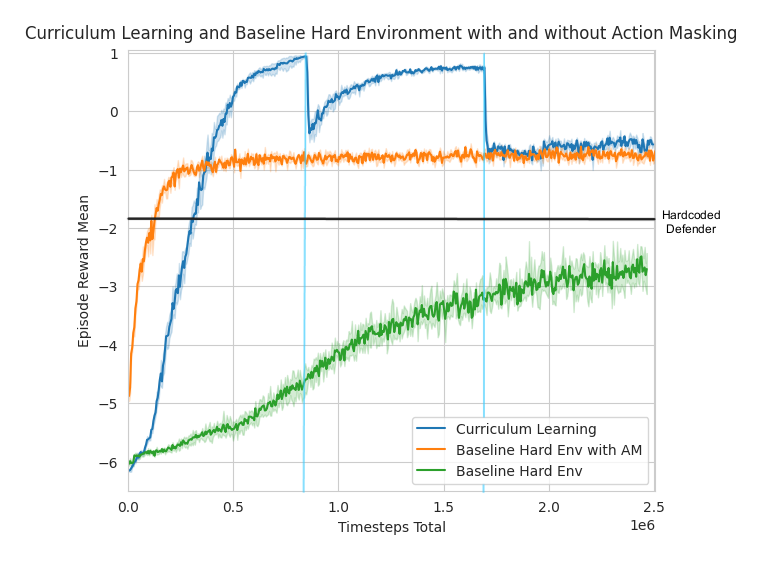}
    \caption{Baseline Hard Environment with and without Action Masking. The changes in task are outlined by the light blue lines for the CL experiments.}
    \label{fig:cl_am}
\end{figure}

In the hard environment configuration with an action mask, the agent reached a higher level of performance than solely training on the hard environment, as shown in Figure \ref{fig:cl_am}, with an episode reward mean of -0.743 in comparison to the baseline hard environment episode reward mean of -2.791. The agent trained with an action mask achieved a slightly lower mean episode reward mean than CL, but was able to reach that episode reward mean at a sharper rate of learning. The AM agent, similarly, to the CL agent, significantly outperformed the hardcoded agent in the hard environment.\par
\newpage

\subsection{Action Masking and Curriculum Learning}
\begin{figure}[hbt!]
    \centering
    \includegraphics[width=0.75\linewidth]{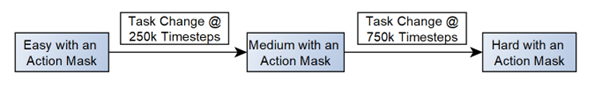}
    \caption{Curriculum for AM and CL Experiment.}
    \label{fig:cl_am_experiment}
\end{figure}

Following the conclusions of the previous sections in this paper AM and CL were applied together with the subsequent curriculum shown in Figure \ref{fig:cl_am_experiment}. \par
The curriculum outlined in Figure 8 shows that the tasks are switched much earlier than the curriculum for the experiment without action masking presented in Figure \ref{fig:cl_experiment}. This is because, as shown in Figure \ref{fig:AM_baseline}, the learning of policies trained with action masking plateaued far sooner than policies trained without action masking. The curriculum is consequently adapted to match the attributes of agents trained with action masking, changing tasks at 250k and 750k timesteps. \par
\newpage
\begin{figure}[hbt!]
    \centering
    \includegraphics[width=0.75\linewidth]{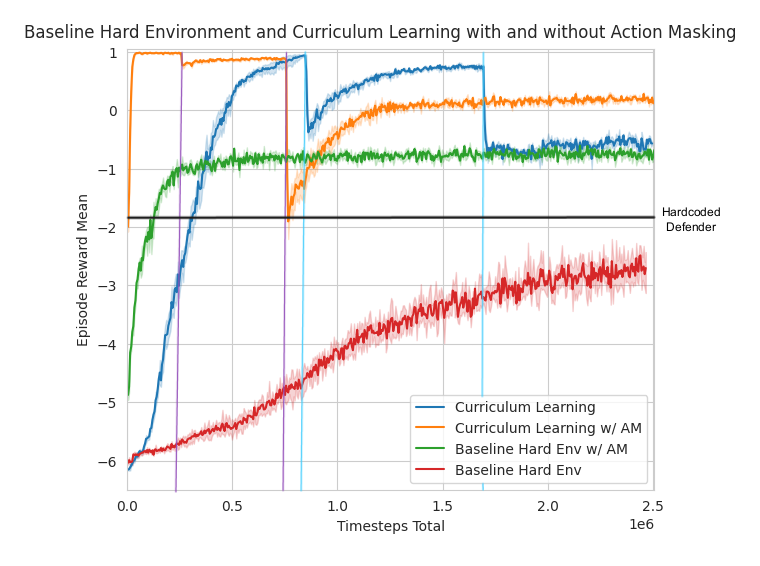}
    \caption{Baseline Hard Environment and Curriculum Learning with and without Action Masking. The changes in task are outlined by the purple lines at the respective timestep totals for AM and CL experiments and by light blue lines for the CL experiments.}
    \label{fig:cl_am_final}
\end{figure} 
In Figure \ref{fig:cl_am_final} it can be observed that when the techniques of CL and AM are combined, for all seeds, an agent trained in these conditions reached a higher mean episode reward of 0.137, in fewer timesteps, than agents trained solely with CL, AM or without either. The combination of CL and AM was therefore shown to produce the highest level of performance in comparison to the other training techniques tested and the hardcoded defender. \par

\section{Conclusion}
This paper demonstrates that the application of action masking and curriculum learning individually improved the overall performance of training a defensive agent to remediate attacks in more complex IPMSRL environment configurations. This includes real work dynamics such as false positive and negative alerts and the delay that is inherent in OT systems. The benefits of applying these techniques were even more pronounced when they were applied together.\par
Curriculum learning alone was shown to increase episode reward mean from -2.791 to -0.569. Action masking alone is shown to similarly improve mean episode reward. In this paper an improvement to -0.743 was seen. \par
Finally, both curriculum learning and action masking are applied together. This training method resulted in the highest level of mean episode reward observed in this paper, with a mean episode reward of 0.137. \par
As the complexity of the environment increased, the hardcoded defender struggled to maintain a strong level of performance. The hardcoded defender was able to outperform vanilla PPO, but the application of AM or CL training techniques enabled the RL defensive agent to achieve a significantly higher episode reward mean than the hardcoded defender in the hard difficulty environment. \par
A potential reason that the hardcoded defender struggled in the hard environment, in comparison to the best performing RL agents, is the uncertainty that a high proportion of FP alerts provides. The hardcoded agent struggles to prioritise the remediation of true alerts as there is no mechanism in its logic to establish whether an alert is a FP or not. It will have to randomly choose between which alert to remediate if there are multiple alerts active. The only prioritisation that is baked into the logic of the hardcoded defender is to focus on nodes adjacent to critical nodes. The RL agents on the other hand may have developed policies that allow them to more efficiently decide which alerts to prioritise. This behaviour is very difficult and time consuming to encode into a hardcoded defender’s logic, further displaying the benefits of using an RL based approach for autonomous cyber security.\par
An important note about the use of curriculum learning is the significance of defining appropriate task changing criteria. This paper implemented a simple curriculum, changing tasks at a specified number of total timesteps. There is further scope to optimise this process and potentially see higher levels of performance.\par
Additionally, there is a trade-off present when using action masking; the masking conditions used during training need to be present when querying the trained policy. This is a drawback in the sense that it adds a dependency to the agent’s deployment, and additional bias is added through the setting of masking conditions. But the benefit of constraining certain actions which don’t meet the requirements set out when developing the masking conditions is a tangible one. The use of action masking in this way therefore benefits from gains in data efficiency, overall performance, and an ability to restrict actions to meet the user-defined requirements of a given system. In future work the use of action masking could begin to consider the safety-critical nature of OT systems. Action masking could further be used to help to build trust in autonomous agents that aim to be applied to real systems.

%%
%% The acknowledgments section is defined using the "acknowledgments" environment
%% (and NOT an unnumbered section). This ensures the proper
%% identification of the section in the article metadata, and the
%% consistent spelling of the heading.
\begin{acknowledgments}
    Research funded by Frazer-Nash Consultancy Ltd. on behalf of the Defence Science and Technology Laboratory (Dstl) which is an executive agency of the UK Ministry of Defence providing world class expertise and delivering cutting-edge science and technology for the benefit of the nation and allies. The research supports the Autonomous Resilient Cyber Defence (ARCD) project within the Dstl Cyber Defence Enhancement programme.

    The authors would also like to thank Lisa Gralewski, Marco Casassa Mont, David Foster, Clare Jubb, Laura Caddy, Tasha Hughes and Jake Rigby for their wider contribution to the project and paper.
\end{acknowledgments}

%%
%% Define the bibliography file to be used
\bibliography{references}
\newpage
%%
%% If your work has an appendix, this is the place to put it.
\appendix
\section{PPO Model Hyperparameters}
\label{Appendix}
\begin{table}[hbt!]
%\centering
% \caption{PPO Model Hyperparameters}
\label{table:hyper} 
% \catcode`,=\active
% \def,{\char`,\allowbreak}
\renewcommand\arraystretch{1.2}
% \begin{tabular}{p{2.4cm}<{\raggedright} p{2cm}<{\raggedright}}
% \centering
\begin{tabular}{p{2.4cm} p{2cm}}
  \toprule
    \textbf{Hyperparameters} & \textbf{Values} \\ 
  \midrule
    fc net activation & Swish \\
    fc net hiddens & [256, 256] \\
    value net hiddens & [32] \\
    lambda & 0.925 \\
    kl coeff & 0.1 \\
    vf clip param & 25 \\ 
    sgd minibatch size & 100 \\
    num sgd iter & 15 \\
    vf loss coeff & 0.75 \\
    entropy coeff & 0.0001 \\
    clip param & 0.2 \\
    lr & 0.0005 \\ 
    gamma & 0.995  \\
    train batch size & 5000 \\
    total timesteps & 2,500,000 \\
  \bottomrule
\end{tabular} 
\end{table}

%%
%% End of file
\end{document}